\documentclass{iau}
\usepackage{graphicx}

\title[Euclid Space Mission: building the sky survey] %% give here short title %%
{Euclid Space Mission: building \\the sky survey}

\author[Tereno, I., Carvalho, C.S., Dinis, J., Scaramella, R. \etal]   %% give here short author list %%
{I. Tereno$^{1,2}$, C.S. Carvalho$^1$, J. Dinis$^{1,2}$, R. Scaramella$^3$, J. Amiaux$^4$, C. Burigana$^5$, J.C. Cuillandre$^4$, A. da Silva$^{1,2}$, A. Derosa$^5$, E. Maiorano$^5$, M. Maris$^6$, D. Oliveira$^7$, P. Franzetti$^8$, B. Garilli$^8$, P. Gomez-Alvarez$^9$, M. Meneghetti$^{10}$, S. Wachter$^{11}$ \and the Euclid Collaboration}
\affiliation{
$^1$Instituto de Astrof\'{i}sica e Ci\^{e}ncias do Espa\c{c}o, Tapada da Ajuda, 1349-018 Lisboa, Portugal, email: {\tt tereno@fc.ul.pt} ,\\
$^2$Faculdade de Ci\^{e}ncias da Universidade de Lisboa, Campo Grande, 1749-016 Lisboa, Portugal\\
$^3$INAF - Osservatorio Astronomico di Roma, Via Frascati 33, 00040 MontePorzio Catone, Italy\\
$^4$Commissariat à l'Energie Atomique, Orme des Merisiers, 91191 Gif sur Yvette, France\\
$^5$INAF - IASF Bologna,  via Piero Gobetti 101, 40129 Bologna, Italy\\
$^6$INAF - Osservatorio Astronomico di Trieste Via G. B. Tiepolo 11, 34131 Trieste, Italy\\
$^7$Max Planck Institut fuer Astrophysik, Karl Schwarzschild Str.1., 85740 Garching, Germany\\
$^8$INAF - IASF Milano, Via E Bassini 15, 20133 Milano, Italy\\
$^9$European Space Agency, ESAC, POB 78, 28691 Villanueva de la Canada, Madrid, Spain\\
$^{10}$INAF - Osservatorio Astronomico di Bologna, Via Ranzani 1, 40127 Bologna, Italy\\
$^{11}$Max Planck Institut fuer Astronomie, Konigstuhl 17, 69117 Heidelberg, Germany
}

\pubyear{2014}
\volume{306}  %% insert here IAU Symposium No.
\pagerange{119--126}
\jname{Statistical Challenges in 21$^{st}$ century Cosmology}
\editors{A.C. Editor, B.D. Editor \& C.E. Editor, eds.}
\begin{document}

\maketitle

\begin{abstract}
The Euclid space mission proposes to survey 15000 square degrees of the extragalactic sky during 6 years, with a step-and-stare technique. The scheduling of observation sequences is driven by the primary scientific objectives, spacecraft constraints, calibration requirements and physical properties of the sky. We present the current reference implementation of the Euclid survey and on-going work on survey optimization.
\end{abstract}

\firstsection % if your document starts with a section,
\section{Introduction}

The Euclid space mission (\cite[Laureijs et al 2011]{redbook}), part of the European Space Agency Cosmic Vision Program, is currently under construction by a large international consortium of research centres together with the ESA and aerospace industry partners. The mission addresses the big scientific questions related to fundamental physics and cosmology on the nature and properties of dark energy, dark matter and gravity, as well as on the physics of the early universe and the initial conditions that seed the formation of cosmic structure. For this purpose, the collaboration is building a space telescope with associated visible imager, near infrared photometer and slitless spectrograph, scheduled to be launched on 2020 and to stay in operation for 6 years. Euclid will map the sky with a step-and-stare strategy imaging an average of 30 galaxies per square arc-minute, with mean redshift around 1, which means a total of around 2 billion galaxies over 15000 square degrees of the extragalactic sky, and obtaining the spectra of an average of 3500 galaxies for square degree. The large catalogue will be used to detect signatures of the expansion rate of the Universe and the growth of cosmic structures using two main probes: weak gravitational lensing effects on galaxies and the properties of galaxy clustering.

The mission is defined as a survey that aims at tilling the useful sky with the reference elementary observation pattern in the most efficient way. The elementary observation pattern includes a dithering pattern of 4 frames and the exposure time (including exposure time per instrument and readout overheads). The survey consists of a Wide Survey that ensures the performance of the primary science and a Deep Survey that allows for legacy science. A large volume of calibration data, amounting to $20\%$ of the total observing time, is needed for characterisation of the on-board instruments and for sample characterization in order to control systematic effects. These data are obtained in various time scales, using both survey data and dedicated observations on specific targets.

\section{The Reference survey}
The limits of the extragalactic sky that meet the area and depth requirements needed for the two main probes to reach the science objectives, are set by considering the main contributors to galaxy counts. These are zodiacal background and extinction, which impact the signal-to-noise ratio (SNR), and bright stars, which impact the number of useful objects.
The optimal region excludes low galactic latitudes and low ecliptic latitudes.

The Wide survey is built on the optimal region at the time slots not occupied by calibrations. Indeed, given the needs for specific cadences and periodic observations, calibration observations are defined first, fixing a time skeleton. The result of the galaxy counts analysis also helps in defining the strategy. The survey is built by starting observations at the region of higher density of galaxies, starting from the ecliptic poles where the zodiacal background is minimum, so to have best SNR in the early stages of the mission. The Deep survey is built on fields located close to the ecliptic poles to maximize visibility from space. The basic survey strategy is implemented using the Euclid Sky Survey Planning Tool and described in \cite[Amiaux et al 2012]{refsurvey12}.

When building the survey, stringent constraints on the spacecraft attitude must be followed. Indeed, in order to ensure that the thermo-mechanical stability of the Payload Module is sufficient to preserve the weak lensing measurement accuracy, the spacecraft will be operated with its sunshield normal to the sun direction, with only small variations around this attitude being authorized. The telescope is only allowed to depoint from the orthogonal direction to the centre of the solar disk by a solar aspect angle of -3 to 10 degrees. If observing with a small depointing angle, the field-of-view on the sky would be tilted. To minimize gaps in the sky coverage, this tilt may be compensated by observing with an equivalent tilt in the telescope. The authorized range of this tilt is severely constrained by the structure of the solar shield. Other important limits to take into account are the total authorized number of telescope operations and the maximum allowed number of large slews between non spatial consecutive pointings.

\begin{figure}[h]
% \vspace*{-2.0 cm}
\begin{center}
 \includegraphics[width=12cm]{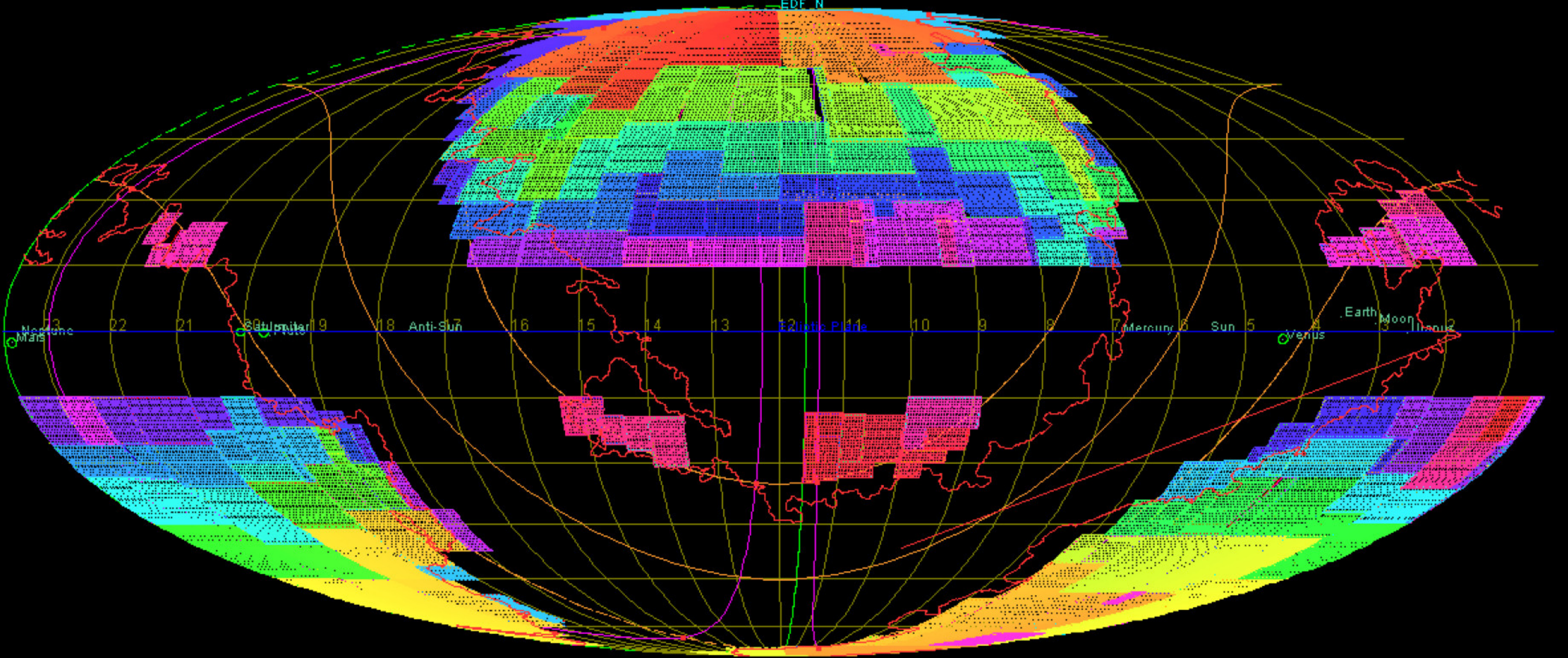} 
% \vspace*{-1.0 cm}
 \caption{The present Euclid reference survey fulfilling all mission specifications, shown in a Mollweide projection of the entire sky in ecliptic coordinates with the ecliptic North pole up.  Jagged line is an iso-contour of extinction. Different colours indicate different years of the survey.}
   \label{fig1}
\end{center}
\end{figure}

Given all these science and technical constraints, we have produced a survey, which is the current reference to be used in all relevant Euclid applications, and is shown in Fig.\,\ref{fig1}. This reference survey achieves a sky coverage of 15000 square degrees in 5.5 years, with over 40000 pointings all within the depointing constraints and $85\%$ with small tilts (less than $1\deg$). Calibrations are made in time blocks of up to one week, and the number of large slews is kept under 10 per month on average.

\section{Optimizing the Euclid Survey}

We aim to produce various alternative versions of compliant surveys, which calls for a larger degree of automation. On the other hand, the current reference survey has room for optimization. To meet these needs, we present two complementary approaches: a ‘deterministic’ one, where the sequence is constructed step by step, and a ‘stochastic’ one, where we seek for a global solution. In both cases we start by pre-tessellating the sky with around 50000 identical fields (corresponding to the telescope field-of-view, FoV).

 {\underline{\it Computing the Euclid Survey from a Set of Conditions}}.  At a given time, only a small part of the sky is available for observation due to the depointing constraints, defining a spatial time window. We start by considering all FoVs within the time window and build one sequence of FoVs for the time slot between two consecutive calibration blocks. From a starting field at high latitude, we build a progression mostly up and down ecliptic meridians, within a limited latitude range producing thus horizontal bands. The progression should leave no gaps, minimize the number of large slews and prioritize higher quality fields. There are fixed conditions to determine the latitude range of the horizontal bands, to recover missing fields in a compact region (left during calibration times) and to establish when and where to shift between hemispheres. 
The conditions may also be used as variables in a stochastic approach.

 {\underline{\it Computing the Euclid Survey with Simulated Annealing}}. We consider the Euclid survey as a combinatorial optimization problem, whose solution is a sequence that visits all the fields without repetitions, in the shortest time, and fulfils all constraints. In this approach, the space of configurations (all possible sequences) will be searched in a controlled manner. We again consider the time windows, but this time they are defined taking into account both depointing and tilting constraints. For each time slot between two consecutive calibration blocks, we define a number of patches grouping neighbouring fields and tile them in an optimal way. The tiled patches for all time windows form an initial sequence. We generate neighbouring sequences from a current one based on random sets of changes. We will use the simulated annealing technique (\cite[Kirkpatrick et al 1983]{sa}) to evaluate the sequences. A sequence with lower cost or with a slight increase in cost (to escape from local minima) is kept, in a converging process. Gradually the acceptance threshold (the so-called temperature parameter) is lowered, restricting the acceptance of new sequences and ensuring convergence to a sequence with minimal idle time.

\end{document}